\documentclass[aps,prd,preprint,superscriptaddress,showpacs,nofootinbib]{revtex4}

\usepackage{graphicx}

\begin{document}

\preprint{UTAP-426} \preprint{astro-ph/0210502}

\title{Supernova relic neutrinos and observational implications for
neutrino oscillation}


\author{Shin'ichiro Ando}
\email[Email address: ]{ando@utap.phys.s.u-tokyo.ac.jp}
\affiliation{Department of Physics, School of Science, the University of
Tokyo, 7-3-1 Hongo, Bunkyo-ku, Tokyo 113-0033, Japan}

\author{Katsuhiko Sato}
\affiliation{Department of Physics, School of Science, the University of
Tokyo, 7-3-1 Hongo, Bunkyo-ku, Tokyo 113-0033, Japan}
\affiliation{Research Center for the Early Universe, School of Science,
the University of Tokyo, 7-3-1 Hongo, Bunkyo-ku, Tokyo 113-0033, Japan}


\date{\today}

\begin{abstract}
We investigate the flux of supernova relic neutrinos (SRN) for several
 neutrino oscillation models with parameters inferred from recent
 experimental results.
In the calculation, we adopt the realistic {\it time-dependent}
 supernova density profile, which is very different from the static
 progenitor profile owing to shock propagation.
The Earth matter effect is also included appropriately using realistic
 density profile of the Earth.
As a result, these two effects are found to induce the flux difference
 by a few \% in the detection energy range ($E_\nu > 19.3$ MeV).
We also set 90\% C.L. upper limit on SRN flux for various oscillation
 models using the recently released observational result by
 Super-Kamiokande (SK).
In the near future, further reduced upper limit is actually expected,
 when the current SK data are reanalyzed using some technique to reduce
 background events against SRN signals.
It is expected that the reduced upper limit is sufficient to provide
 useful implications for neutrino oscillation.
\end{abstract}


\maketitle

\section{Introduction \label{sec:Introduction}}

A core-collapse supernova explosion produces a number of neutrinos and
99\% of the gravitational energy is transformed to neutrinos.
It is generally believed that the core-collapse supernova explosions
have traced the star formation history in the universe and have emitted
a great number of neutrinos, which should make a diffuse background.
This supernova relic neutrino (SRN) background is one of the targets of
the currently working large neutrino detectors such as Super-Kamiokande
(SK).
Comparing the predicted SRN flux with the observations provides
potentially valuable information on the nature of neutrinos as well as
the star formation history in the universe.
This SRN background has been theoretically discussed in a number of
previous papers \cite{Totani95,Ando02}.

On the other hand, there are observational constraints on the SRN flux.
Most recently, SK collaboration set an upper bound of 1.2 $\bar{\nu}_e
~\mathrm{cm^{-2}~s^{-1}}$ for the SRN flux in the energy region $E_{\nu}
> 19.3 ~\mathrm{MeV}$ \cite{Malek02}.
(This limit can constrain only the flux of $\bar{\nu}_e$, since that
flavor is the most easily detected by SK.)
It is two orders of magnitude lower than the previous one obtained by
Kamiokande II \cite{Zhang88}, and is the same order as some typical
theoretical predictions \cite{Totani95,Ando02}.
For example, Ando et al. \cite{Ando02} (hereafter AST) predicted that
the total SRN flux integrated over entire energy was
$11~\mathrm{cm^{-2}~s^{-1}}$, while the corresponding SK limit
calculated with the AST spectral shape is $31 ~\mathrm{cm^{-2}
~s^{-1}}$.
Since the theoretical calculations contain many ambiguities such as the
supernova rate in the universe and neutrino spectrum from each
supernova, this severe observational SRN limit can provide a number of
valuable information on the various fields of astrophysics and
cosmology (e.g. Ref. \cite{Fukugita02}).
Further, in the near future, it is expected that the upper limit will be
much lower (about factor 3) when the current SK data of 1,496 days are
reanalyzed using some technique to reduce background against detection.
In that case, the SRN signal might be detected and the very severe
constraint not only on the star formation history but also on the nature
of neutrinos might be obtained.

Thus, it is obviously important and very urgent to give more precise
prediction for the SRN flux and event rate.
For that reason, in this paper, we investigate the SRN flux using the
most realistic models to date.
In particular, we include the new features which have not been
considered in all of the past studies \cite{Totani95,Ando02},
illustrating them below.
First, when we calculate the neutrino conversion probability in
supernova, we adopt the realistic {\it time-dependent} density and $Y_e$
profiles, which are calculated by the Lawrence Livermore group
\cite{Takahashi02b}.
During the neutrino burst ($\sim 10$ sec), the shock wave propagating
the supernova matter changes density profile dramatically, and it is
expected to affect the adiabaticity of resonance points
\cite{Schirato02}.
Second, we consider the Earth matter effect.
(For that effect on the future Galactic supernova neutrino burst, see
Refs. \cite{Dighe00,Lunardini01,Takahashi02} and references therein.)
Since SRN come from all the directions, half of them pass through the
Earth matter, and it is also expected to change the SRN spectrum.
Finally, neutrino oscillation with inverted mass hierarchy $(m_3 \ll
m_1)$ is also investigated.
In that case, the resonance also occurs in anti-neutrino sector, and it
is expected that the SRN spectrum would be quite different from that in
the case of normal mass hierarchy.
Further, we repeat the discussions given in the recent SK paper
\cite{Malek02} and obtain the 90\% C.L. upper limit for the SRN flux
estimated using various oscillation models.

This paper is organized as follows.
In Section \ref{sec:Neutrino conversions and supernova models}, we give
the models of original neutrino spectrum and supernova density profile. 
We also illustrate the neutrino oscillation models investigated in this
paper, and qualitative behavior of flavor conversions for these models.
In Section \ref{sec:Results}, we show the calculated SRN fluxes, and in 
Section \ref{sec:Discussion}, we compare the result with AST
calculation. 
We also compare them to the observational upper limit from SK and
discuss the validity of various oscillation models.
Model dependence of our calculations are also discussed in the same
section.


\section{Adopted models and neutrino conversions
\label{sec:Neutrino conversions and supernova models}}

\subsection{Supernova and Earth models \label{sec:Supernova models}}

As for the neutrino spectrum at production, we use that calculated by
the Lawrence Livermore group \cite{Totani98}.
The mean energies are different between flavors, such as
\begin{equation}
\langle E_{\bar{\nu}_e} \rangle \simeq 16 ~\mathrm{MeV}, 
~\langle E_{\bar{\nu}_x} \rangle \simeq 22 ~\mathrm{MeV}.
\label{eq:average energy}
\end{equation}
This hierarchy of mean energies is explained as follows.
Since $\bar{\nu}_x$'s interact with matter only through the
neutral-current reactions in supernova, they are weakly coupled with
matter compared to $\bar{\nu}_e$'s.
Thus the neutrino sphere of $\bar{\nu}_x$'s is deeper in the core than
that of $\bar{\nu}_e$'s, which leads to higher temperatures for
$\bar{\nu}_x$'s.
Therefore, flavor conversions are expected to enhance the mean energy of
$\bar{\nu}_e$'s.
However, recent studies of neutrino flux and spectra formation in a
supernova core (e.g. Ref. \cite{Keil02}) have shown that average
$\bar{\nu}_x$ energy exceeds the average $\bar{\nu}_e$ energy by only a
small amount, 10\% being a typical number.
We discuss this new aspect in Section \ref{sec:Uncertainty concerning
supernova neutrino spectra}.

We calculate neutrino conversion probabilities in supernova using the
realistic {\it time-dependent} density and $Y_e$ profiles, which are
also calculated by the Lawrence Livermore group \cite{Takahashi02b}.
Among the past studies \cite{Totani95,Ando02}, AST first estimated the
effects of neutrino oscillation quantitatively, however, they calculated
conversion probability using static pre-collapse progenitor model 
\cite{Woosley95}.
As the recent study indicates \cite{Schirato02}, the density profile
changes drastically during neutrino burst ($\sim 10$ sec) owing to
shock propagation in supernova matter.
When the shock propagates through the regions where matter-enhanced
neutrino flavor conversion occurs (so called resonance point), it is
expected to affect the adiabaticity of the resonance.
(For the resonance and its adiabaticity, see the next subsection for
details.)
We show the $\rho Y_e$ profile of supernova at 0.5, 2, 5, 10, and 15
seconds after bounce in the upper panel of Fig. \ref{fig:shock}.
As is seen clearly, the resonance takes place more than once at $\agt 2$
sec, and this is not expected when we use the static progenitor model.
(As is discussed in the next subsection, the resonance for the
anti-neutrino sector occurs only in the case of inverted mass hierarchy,
or INV model in our notation.)

Further, we include the Earth matter effect using realistic density
profile of the Earth \cite{Dziewonski81}.
This effect has not been considered in the past studies including AST
\cite{Totani95,Ando02}, on the other hand, it has been discussed in the
case of the future Galactic supernova neutrino burst by various authors
(see Refs. \cite{Dighe00,Lunardini01,Takahashi02} and references
therein).
Although the effect is expected to be very small for anti-neutrinos 
\cite{Dighe00,Lunardini01,Takahashi02}, we include it because the half
of the SRN pass through the Earth matter.

\subsection{Neutrino conversions \label{sec:Neutrino conversions}}

As for oscillation parameters, we adopt the latest results which are
updated recently by the SK and SNO solar neutrino observations
\cite{Fukuda02,Ahmad02}.
We adopt only large mixing angle (LMA) solution, whose parameters are
\begin{equation}
\Delta m^2_{12} = 5.0 \times 10^{-5} ~\mathrm{eV^2},
~ \tan^2 \theta_{12} = 0.42.
\end{equation}
On the other hand, from the atmospheric neutrino experiments
\cite{Fukuda99}, we use the values
\begin{equation}
|\Delta m^2_{13}| = 2.8 \times 10^{-3} ~\mathrm{eV^2},
~ \sin^2 2\theta_{23} = 1.0,
\end{equation}
in our calculations.
Whether $\Delta m^2_{13} = m^2_3 - m^2_1$ is positive (normal mass
hierarchy) or negative (inverted mass hierarchy) is not known well,
although the future Galactic neutrino burst is expected to provide
useful information on the mass hierarchy \cite{Dighe00,Lunardini01,
Takahashi02c} and actually several implications for it have been
obtained with the neutrino burst from SN 1987A \cite{Lunardini01b}.
For $\theta_{13}$, which is not also sufficiently constrained, we adopt
two large and small values,
\begin{equation}
\sin^2 2\theta_{13} = 0.04, ~ 1.0 \times 10^{-6},
\end{equation}
both of which satisfy the upper bound from reactor experiment
\cite{Apollonio99}.
(Investigating neutrino oscillation using intermediate values for
$\theta_{13}$ is beyond the scope of this study, however, it is given in
Ref. \cite{Takahashi02b} for the future Galactic supernova neutrino
burst.)
From these discussions, we adopt four parameter sets, named as NOR-S,
NOR-L, INV-S, and INV-L, where NOR and INV represent the normal and
inverted mass hierarchy respectively.
The suffixes -L and -S attached to NOR and INV stand for large and small
$\theta_{13}$, respectively.
In addition, we also investigate in case of no oscillation, for
comparison.

Now, we qualitatively discuss the behavior of anti-neutrino conversions
during propagation in supernova matter\footnote{Here, we note that there
is a wrong statement in Section 2.2 in AST. We wrote there ``We can
naively deal with anti-neutrino oscillation effect as vacuum oscillation,
since $\bar{\nu}_e$'s are not affected by the resonance.'' As is
illustrated in the text, this is wrong, however, the calculations in AST
is including supernova matter effect appropriately.} (see
e.g., Ref. \cite{Dighe00} for details).

We consider, first, the case of normal mass hierarchy.
The state of $\bar{\nu}_e$ produced at deep in the core is coincide with
mass eigenstate $\bar{\nu}_1$, owing to large matter potential.
This state can propagate to the supernova surface without being
disturbed by the level crossing between different mass eigenstates (it
is said that there are no resonance points).
Thus, $\bar{\nu}_e$ at production becomes $\bar{\nu}_1$ at the surface
of the supernova and the observed $\bar{\nu}_e$ flux is given by
\begin{eqnarray}
F_{\bar{\nu}_e} &=& |U_{e1}|^2F_{\bar{\nu}_1} + |U_{e2}|^2F_{\bar{\nu}_2}
 + |U_{e3}|^2F_{\bar{\nu}_3} \nonumber \\ 
 &=& |U_{e1}|^2F^0_{\bar{\nu}_e} + (1-|U_{e1}|^2)F^0_{\bar{\nu}_x},
\label{eq:flux_normal}
\end{eqnarray}
where $U_{\alpha i}$ is the mixing matrix between mass and flavor
eigenstates, $F$ the flux at the Earth, $F^0$ the flux at production,
and the suffix $x$ represents $\mu$ and $\tau$, of which flavor
neutrinos are considered to have the same flux.
As $|U_{e1}|^2\sim 0.7$ for the LMA models, the flavor mixing is
expected to harden the $\bar\nu_e$ spectrum.

\begin{figure}[htbp]
\includegraphics[width=15cm]{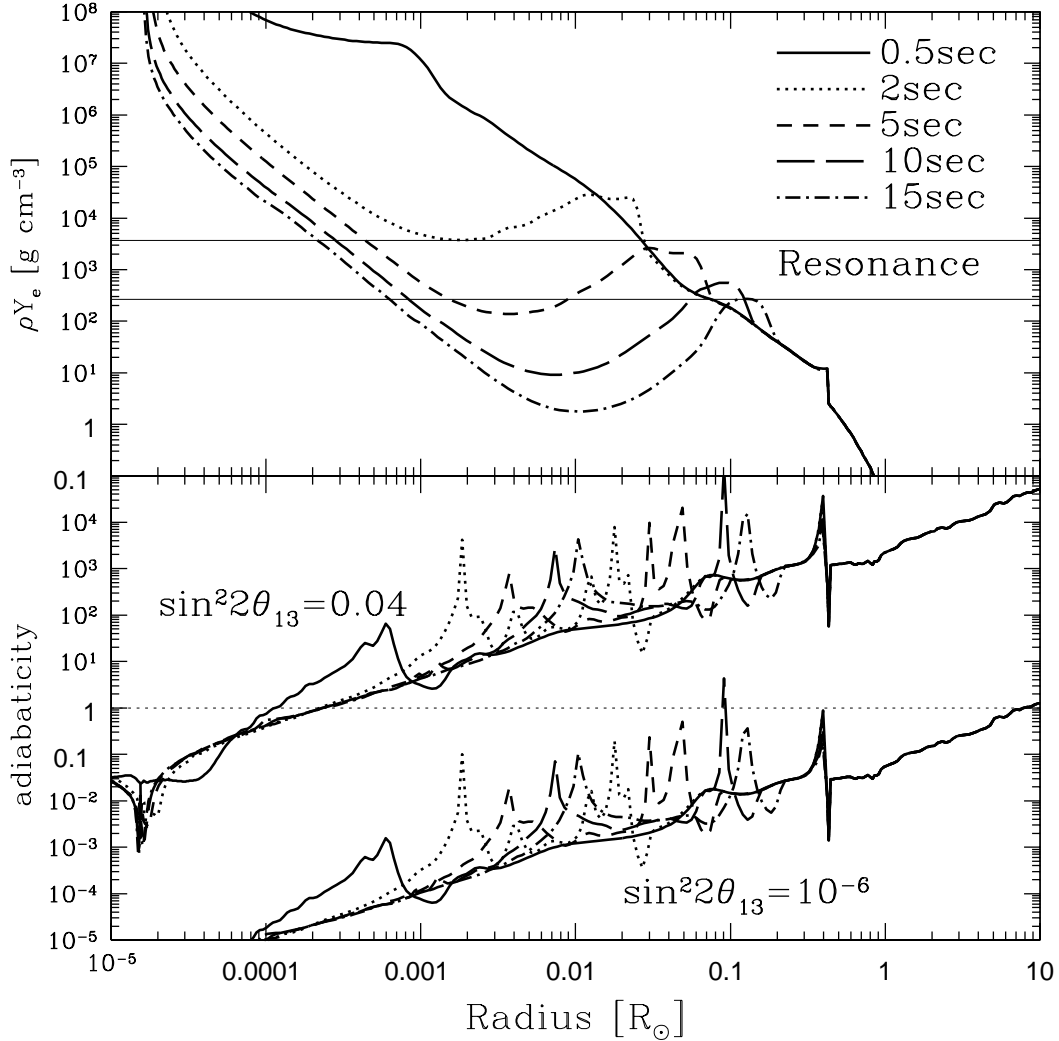}
\caption{Upper panel: The $\rho Y_e$ profile of supernova at 0.5, 2, 5,
 10, and 15 seconds after bounce. The horizontal band shows resonance
 condition for two INV models (the band width comes from the energy
 range 5-70 MeV), i.e., at intersections between the $\rho Y_e$ curve
 and the horizontal band, the MSW resonance occurs. Lower panel: The
 adiabaticity of the resonance (i.e., if the adiabaticity at the
 resonance point is larger than 1, then the resonance is adiabatic) for
 INV-L (labeled by $\sin^22\theta_{13}=0.04$) and INV-S
 ($\sin^22\theta_{13}=10^{-6}$) models. The neutrino energy is assumed
 to be 20 MeV, and the line types are the same as those used in the
 upper panel. \label{fig:shock}}
\end{figure}

However, in the case of inverted mass hierarchy, this situation changes
dramatically.
Since $\bar{\nu}_3$ is the lightest, $\bar{\nu}_e$ is created as
$\bar{\nu}_3$ in the supernova core.
In that case, it is well known that at a so called resonance point,
there occurs a level crossing between $\bar{\nu}_1$ and $\bar{\nu}_3$.
At this resonance point, complete $\bar{\nu}_1 \leftrightarrow
\bar{\nu}_3$ conversion occurs when the so called adiabaticity is
sufficiently small compared to one (it is said that resonance is
nonadiabatic), while never occurs when it is large (adiabatic
resonance).
In the lower panel of Fig. \ref{fig:shock}, we show the adiabaticity
$\gamma$ of each model, which is written by
\begin{equation}
\gamma =\frac{\Delta m_{13}^2\sin^22\theta_{13}}
 {2E_\nu\cos 2\theta_{13}|d\ln n_e/dr|_{\rm res}},
\label{eq:ad_H}
\end{equation}
where $E_\nu$ is the neutrino energy and $n_e$ is the electron number
density; we have assumed $E_\nu =20$ MeV in the figure.\footnote{Note
that the definition of $\gamma$ [Eq. (\ref{eq:ad_H})] is available only
at the resonance points, although in Fig. \ref{fig:shock} we have shown
it as a function of radius over the entire region in supernovae.}
The behavior of the adiabaticity shown in the lower panel of
Fig. \ref{fig:shock} can be qualitatively understood as follows.
The supernova profile on which the adiabaticity depends is the gradient
of logarithmic density, $|d\ln n_e/dr|^{-1}$.
Then, if we assume the simple power-law density profile ($n_e=n_0
r^{-\alpha}$, $n_0$ and $\alpha$ are constants), the adiabaticity is
proportional to the radius.
This global behavior can be seen in the lower panel of
Fig. \ref{fig:shock}; changes in the power-law index $\alpha$ and $n_0$
are responsible for the variability of $\gamma$ in the small scale.

From Fig. \ref{fig:shock}, we can expect that for INV-S, the resonance
is always nonadiabatic.
On the other hand, for INV-L, it is basically adiabatic but at the first
resonance point, the adiabaticity becomes comparable to one particularly
at $\agt 10$ sec.
The situation is very complicated for INV-L model, and there is no way
but to calculate numerically for the case.
From here, however, we estimate the flux in the case of the resonance is
completely nonadiabatic or completely adiabatic, for simplicity.
When the resonance is nonadiabatic, the situation is the same as in the
case of normal mass hierarchy (because $\bar{\nu}_e$ at production
becomes $\bar{\nu}_1$ at the stellar surface), and the $\bar{\nu}_e$
flux we observe is represented by Eq. (\ref{eq:flux_normal}).
On the other hand, adiabatic resonance forces $\bar{\nu}_e$ at
production to become $\bar{\nu}_3$ when it appears from the stellar
surface and therefore, the observed $\bar{\nu}_e$ flux is given by 
\begin{equation}
F_{\bar{\nu}_e} = |U_{e3}|^2F^0_{\bar{\nu}_e}
 + (1-|U_{e3}|^2)F^0_{\bar{\nu}_x}.
\label{eq:flux_inverted}
\end{equation}
Since $|U_{e3}|^2$ is constrained to be much smaller than 1 from reactor
experiment \cite{Apollonio99}, Eq. (\ref{eq:flux_inverted}) indicates
that complete conversion takes place between $\bar{\nu}_e$ and
$\bar{\nu}_{\mu,\tau}$.

\section{Results \label{sec:Results}}

We calculate the SRN flux using the formula
\begin{equation}
\frac{dF_\nu}{dE_\nu}=c \int_0^{z_{\mathrm{max}} }
 R_{\mathrm{SN}}(z) \frac{dN_\nu(E_\nu ^\prime)}{dE_\nu ^\prime}
 (1+z) \frac{dt}{dz} dz,
\end{equation}
where $E_\nu ^\prime = (1+z)E_\nu$, $R_{\mathrm{SN}}(z)$ is supernova
rate per comoving volume at redshift $z$, $dN_\nu / dE_\nu$ energy
spectrum of emitted neutrinos, $z_{\mathrm{max}}$ the redshift when the
gravitational collapses began, which we assumed to be 5.
As supernova rate, we use the most reasonable model to date, which is
based on the optical/UV observation of star formation history in the
universe by Hubble Space Telescope \cite{Madau96}, and the model was
also used in AST as ``SN1''.
In this model, the supernova rate exponentially increases with $z$ and
peaks at $z \sim 1.5$ and exponentially decreases in further high-$z$
region (see Fig. 1 in Ref. \cite{Porciani01}).
Since for the optical/UV observation, there is still uncertainty due to
dust extinction, we introduce a factor $f$ as $R_{\mathrm{SN}}(z) = f
R_{\mathrm{SN1}}(z)$ and assume that it is independent of $z$.
(This argument is almost the same one as that in
Ref. \cite{Fukugita02}.)
The case $f=1$ corresponds to the local supernova rate of
$R_\mathrm{SN}(0) = 8.5 \times 10^{-5} h_{70}^3 ~\mathrm{yr^{-1}
~Mpc^{-3}}$, while the observational local supernova rate is
$R_\mathrm{SN} = (1.2 \pm 0.4) \times 10^{-5} h_{70}^3 ~\mathrm{yr^{-1}
~Mpc^{-3}}$ \cite{Madau98}, where $h_{70} = H_0 / 70 ~\mathrm{km ~s^{-1}
~Mpc^{-1}}$.
For a while, we set $f=1$ as a fiducial value.

\begin{figure}
\includegraphics[width=15cm]{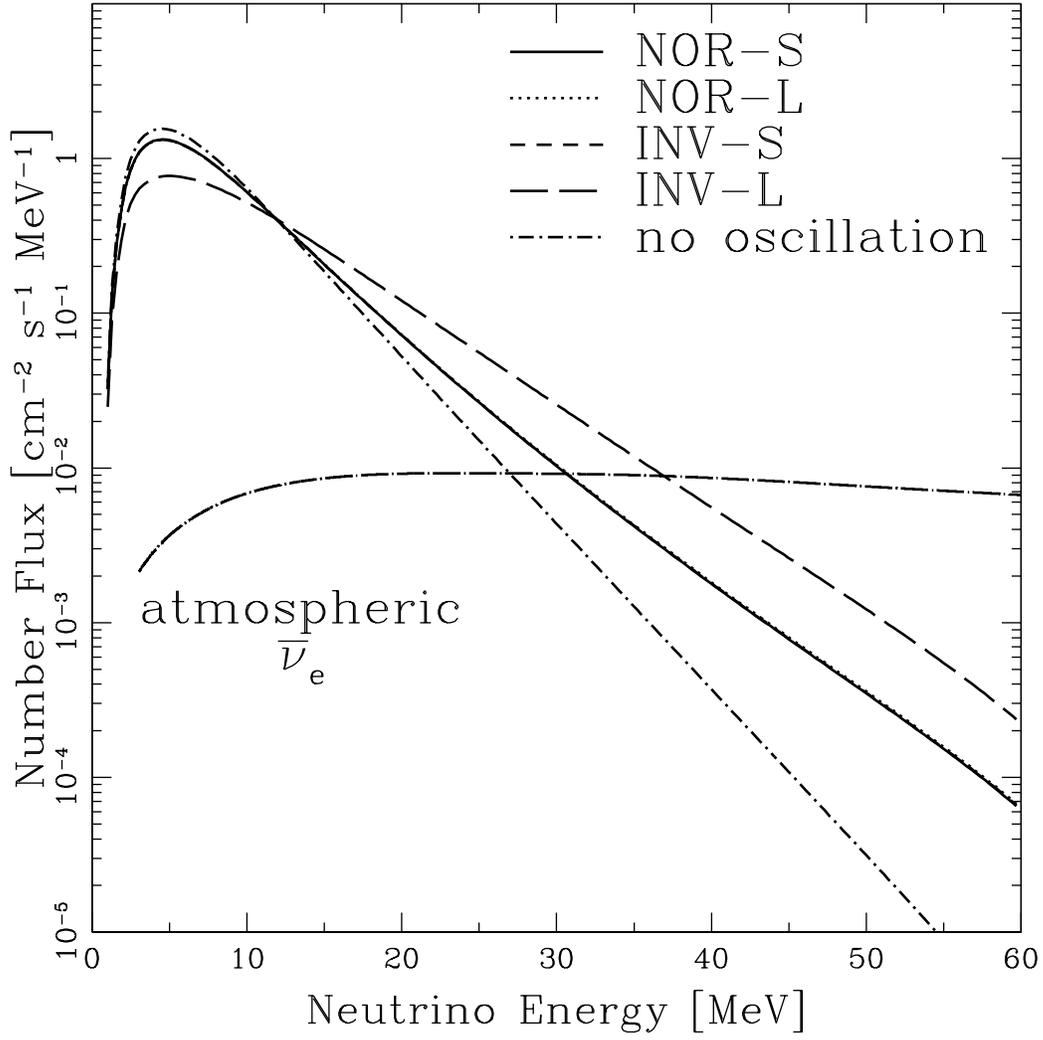}
\caption{Number flux for $\bar{\nu}_e$'s for various neutrino
 oscillation models. The spectra for NOR-S, NOR-L, and INV-S are
 degenerated, while that for INV-L is the hardest one. The flux of
 atmospheric $\bar{\nu}_e$ is also shown. \label{fig:SRN_flux}}
\end{figure}

Figure \ref{fig:SRN_flux} shows SRN flux for the various oscillation
models explained in Section \ref{sec:Neutrino conversions and supernova
models}.
Flavor conversions enhance the average $\bar{\nu}_e$ energy.
Three models NOR-S, NOR-L, and INV-S are degenerated and the model INV-L
has the hardest energy spectrum as is expected from the qualitative
discussions in Section \ref{sec:Neutrino conversions} [see
Eqs. (\ref{eq:flux_normal}) and (\ref{eq:flux_inverted})].
In the same figure, we also show the flux in the case of no oscillation,
for comparison.
In that case, the flavor conversion does not take place, then the
original $\bar\nu_e$ flux is actually observed [$F_{\bar\nu_e}=F_{\bar
\nu_e}^0$, in contrast with the Eqs. (\ref{eq:flux_normal}) and
(\ref{eq:flux_inverted}) for the oscillation cases], resulting in the
softest spectrum in the five models under consideration.

Recent SK upper bound for the SRN flux \cite{Malek02} constrain the
theoretical predictions by factor $\sim 3$.
For example, AST predicted the SRN flux integrated over entire energy to
be $11 ~\mathrm{cm^{-2}~s^{-1}}$, while the corresponding SK limit
calculated with the AST spectral shape is $31 ~\mathrm{cm^{-2}~s^{-1}}$.
The AST model cited above corresponds to NOR-L model in this study.
We repeat the analysis given in the SK paper \cite{Malek02} and obtain
the flux upper limit at 90\% C.L. for various oscillation models.
We summarize the results in Table \ref{table:limit}.
As shown in the table, all the oscillation models are not ruled out yet,
since the theoretical predictions are still smaller than the
corresponding SK limit, while the observational upper limit is rather
severer for the INV-L model.
\begin{table*}[htbp]
\caption{The predicted SRN flux for various oscillation models and the
 corresponding SK limit (90\% C.L.) \cite{Malek02}. The ratio between
 the prediction and the limit is shown in the fourth column. Event rate
 from 19.3 MeV to the energy where the SRN flux dominates the
 atmospheric $\bar{\nu}_e$ flux is also shown in the fifth
 column. \label{table:limit}}
\begin{tabular}{ccccc} \hline \hline
Model & ~Predicted flux & ~SK limit (90\% C.L.) & ~Prediction/Limit &
 ~Event rate for $E_\nu > 19.3$ MeV\\
 \hline
NOR-S & 12 cm$^{-2}$ s$^{-1}$ & $<$ 35 cm$^{-2}$ s$^{-1}$ & 0.34 & 0.80
 /year ($E_\nu < 30$ MeV)\\ 
NOR-L & 11 cm$^{-2}$ s$^{-1}$ & $<$ 34 cm$^{-2}$ s$^{-1}$ & 0.33 & 0.81
 /year ($E_\nu < 30$ MeV)\\
INV-S & 11 cm$^{-2}$ s$^{-1}$ & $<$ 34 cm$^{-2}$ s$^{-1}$ & 0.33 & 0.81
 /year ($E_\nu < 30$ MeV)\\
INV-L & 9.0 cm$^{-2}$ s$^{-1}$ & $<$ 12 cm$^{-2}$ s$^{-1}$ & 0.74 & 2.0
 /year ($E_\nu < 37$ MeV)\\
no oscillation & 12 cm$^{-2}$ s$^{-1}$ & $<$ 73 cm$^{-2}$ s$^{-1}$ & 0.17
 & 0.43 /year ($E_\nu < 27$ MeV)\\ \hline
\end{tabular}
\end{table*}


\section{Discussion \label{sec:Discussion}}

\subsection{Comparison with the other calculations}

\begin{figure}[htbp]
\includegraphics[width=15cm]{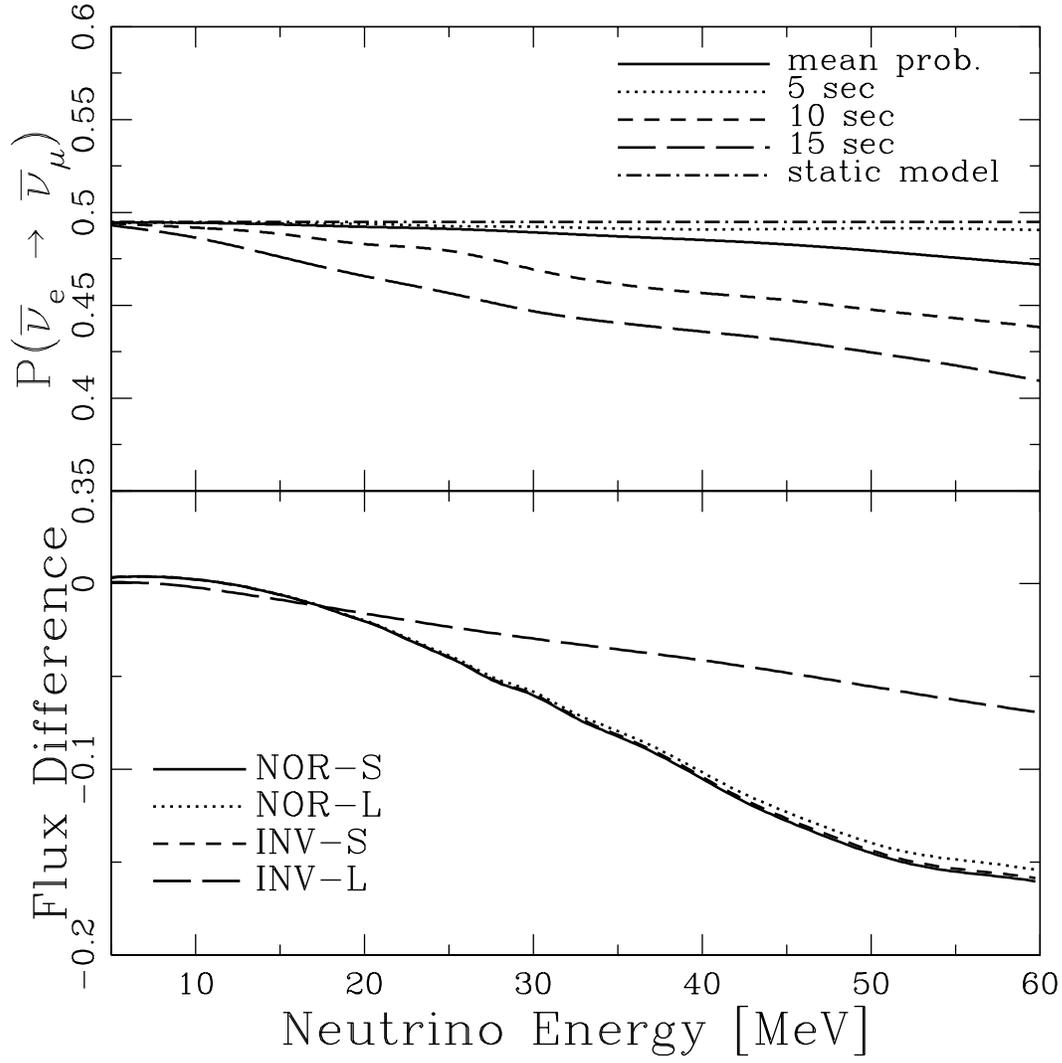}
\caption{Upper panel: Conversion probability $P(\bar{\nu}_e 
 \to \bar{\nu}_\mu)$ as a function of neutrino energy calculated using
 static progenitor model (dot-dashed line), shock model at 5 (dotted
 line), 10 (short-dashed line), and 15 seconds (long-dashed line) after
 bounce. Time-averaged probability by the flux is also shown by
 solid line. Lower panel: Flux difference from the model obtained with
 static progenitor model and without the Earth matter effect, for
 various oscillation models. \label{fig:conversion}}
\end{figure}

The conversion probability for INV-L model is expected to be
time-dependent as discussed in Section \ref{sec:Neutrino conversions}.
We show in the upper panel of Fig. \ref{fig:conversion} the conversion
probability $P(\bar{\nu}_e \to \bar{\nu}_\mu)$ obtained with density
profiles for 5, 10, and 15 seconds after bounce.
The conversion probability in the case that the static progenitor model
is used is also shown for comparison.
At 5 seconds after bounce, it is almost the same for all energy as in
the case of static model, while at 15 seconds, it changes by $\sim 10$\%
at 30 MeV, and by $\sim 20$\% at 60 MeV.
However, since almost all of neutrinos are emitted by 5 seconds after
bounce, the effective conversion probability is expected to be almost
the same as that in the case of static model.
Actually, we calculated the time-averaged probability by the flux and
the result is also shown in the same figure, indicating only $\sim 5$\%
difference even at 60 MeV.

We also compared the flux calculated in Section \ref{sec:Results} with
that obtained by the AST calculation \cite{Ando02}.
The AST calculation adopted the static progenitor model and did not
include the Earth matter effect.
We show in the lower panel in Fig. \ref{fig:conversion}, the flux
difference as a function of neutrino energy.
For the three models NOR-S, NOR-L, and INV-S, the difference comes from
the presence of the Earth matter effect, since these models are not
affected by the time-dependent density profile as illustrated in Section
\ref{sec:Neutrino conversions}.
The Earth matter effect changes flux by $\sim 15$\% at 60 MeV.
For INV-L model, the Earth matter effect is absent (for this reason, see
e.g., Ref. \cite{Dighe00}), and thus, the flux difference comes from the
difference of the supernova model (static or time-dependent).
As shown in the upper panel of Fig. \ref{fig:conversion}, this effect is
small and it makes only $\sim 5$\% difference at 60 MeV.
In Table \ref{table:difference}, we show the difference of the
energy-integrated flux.
In the detection energy range, $E_\nu > 19.3$ MeV, a few \% difference
is expected.
\begin{table}
\caption{Flux difference from the model obtained with static progenitor
 model and without the Earth matter effect. The flux is integrated by
 energy. The second column is for the flux integrated over the entire
 energy range and the third column is over the detection energy range,
 or $E_\nu>19.3$ MeV. The $+~(-)$ sign of each entry represents that the
 flux increased (decreased) by the following
 number. \label{table:difference}}
\begin{tabular}{ccc} \hline \hline
Model & ~Total flux & ~Flux for $E_\nu > 19.3$ MeV \\ \hline
NOR-S & $+$0.015\% & $-$4.0\% \\
NOR-L & $+$0.015\% & $-$3.9\% \\
INV-S & $+$0.015\% & $-$4.0\% \\
INV-L & $-$0.43\% & $-$2.4\% \\ \hline
\end{tabular}
\end{table}

\subsection{Implications for neutrino oscillation and the future}

Theoretical predictions of the SRN flux depend on the factor $f$, which
we have assumed to be one until here, in a proportional manner.
This parameter $f$ is not known well because it concerns dust extinction
of optical/UV photons, and actually even though we take $f=2$, it does
not conflict with the local supernova rate observation. 
[Observational local supernova rate is $R_{\rm SN} = (1.2 \pm 0.4)
\times 10^{-4} ~h_{70}^3 ~\mathrm{yr^{-1}~Mpc^{-3}}$ \cite{Madau98},
while SN1 model in AST predicts $R_{\rm SN1}(0) = 8.5 \times 10^{-5}
~h_{70}^3 ~\mathrm{yr^{-1}~Mpc^{-3}}$.]
However, even if we adopt $f \agt 1.4$, we cannot conclude that the
INV-L model is already ruled out, although from Table \ref{table:limit}
it appears that the predicted flux of the INV-L model becomes larger
than the corresponding upper limit.
This is because the SRN flux contains uncertainty other than the
parameter $f$, which is that concerning the original neutrino spectrum:
in particular, the difference of average energies between $\bar\nu_e$'s
and $\bar\nu_x$'s, about which we give a more detailed discussion in the
next subsection.
Thus, although the current SRN observation includes useful information
on the combined quantity of supernova rate in the universe and neutrino
oscillation parameters, it is difficult to give an explicit solution to
the one specific problem at the present stage.

Now, we consider the future possibility to detect SRN or to set severer
constraint on neutrino models.
The largest background against SRN detection at SK is so called
invisible muon decay products.
This event is illustrated as follows.
The atmospheric neutrinos produce muons by interaction with the nucleons
(both free and bound) in the fiducial volume.
If these muons are produced with energies below \v{C}herenkov radiation
threshold (kinetic energy less than 53 MeV), then they will not be
detected (``invisible muons''), but their decay-produced electrons and
positrons will be.
Since the muon decay signal will mimic the $\bar{\nu}_e p \to e^+ n$
processes in SK, it is difficult to distinguish SRN from these events.
Recent SK limits are obtained by the analysis including this invisible
muon background.
In the near future, however, it will be plausible to distinguish the
invisible muon signals from the SRN signals, using the gamma rays
emitted from nuclei which interacted with atmospheric neutrinos
\cite{Suzuki02}.
Therefore, if we can detect gamma ray events, whose energies are about
5-10 MeV, before invisible muon events by muon life time, we can
subtract them from the candidates of SRN signals.
In that case, the upper limit would be much lower (by factor $\sim$3)
when the current data of 1,496 days are reanalyzed \cite{Suzuki02}, and
the SRN signal might be detected or the more powerful information on the
combined quantities of the supernova rate and the neutrino mixing
parameters would be obtained.
In Table \ref{table:limit}, we also show the expected event rate of the
SRN signal for various models.
The integrated energy ranges are set from 19.3 MeV to the energy where
the SRN flux dominates the atmospheric $\bar{\nu}_e$ flux.
Then, without invisible muon events, the SK data of 1,496 days would be
sufficient to permit the SRN detection and set very severe constraint on
the neutrino mixing parameters. 
In particular, inverted mass hierarchy with large $\theta_{13}$ (INV-L
model) would be ruled out first among five models we have considered.

There is another possibility to enhance the average $\bar{\nu}_e$
energy, or resonant spin-flavor conversions.
This mechanism is induced by interaction between nonzero magnetic
moment of neutrinos and supernova magnetic field.
We investigate this mechanism in another paper \cite{Ando02b} for
details.

\subsection{Uncertainty concerning supernova neutrino spectra
\label{sec:Uncertainty concerning supernova neutrino spectra}}

In this study, we used the original neutrino spectrum calculated by
the Lawrence Livermore group \cite{Totani98}.
Unfortunately, their study as well as the other published full numerical
supernova collapse simulations have not yet included the nucleon
bremsstrahlung process or nucleon recoils, even though it is no longer
controversial that these effects are important.
Recent studies (e.g. Ref. \cite{Keil02}) including all these processes
have shown that average $\bar{\nu}_x$ energy exceeds the average
$\bar{\nu}_e$ energy by only a small amount, 10\% being a typical
number.
If it is the case, the oscillated SRN flux is likely to be close to the
value in the case for no oscillation, and the recent SK limit is not
severe enough to constrain the various oscillation parameters.
However, it is premature to conclude that their results are correct,
since it is based on the neutrino transport study on the background of
an assumed neutron star atmosphere, and this approach lacks
hydro-dynamical self-consistency.
Further, it is also because the mean energies and their ratios change
significantly between the supernova bounce, accretion phase, and the
later neutron star cooling phase.
Whichever is the case, the future SK limit is expected to be
sufficiently severe to constrain various oscillation parameters as well
as supernova rate in the universe.


\begin{acknowledgments}
We would like to thank Super-Kamiokande collaboration including
 Y. Suzuki for useful discussions, and also would like to thank
 H.E. Dalhed for giving numerical data of density profiles including
 shock propagation.
This work was supported in part by grants-in-aid for scientific research
 provided by the Ministry of Education, Science and Culture of Japan
 through Research grant no. S14102004.
\end{acknowledgments}


\begin{thebibliography}{99}

\bibitem{Totani95}
T. Totani and K. Sato,
Astropart. Phys. {\bf 3}, 367 (1995);
T. Totani, K. Sato, and Y. Yoshii,
Astrophys. J. {\bf 460}, 303 (1996);
R.A. Malaney,
Astropart. Phys. {\bf 7}, 125 (1997);
D.H. Hartmann and S.E. Woosley,
Astropart. Phys. {\bf 7}, 137 (1997);
M. Kaplinghat, G. Steigman, and T.P. Walker,
Phys. Rev. D {\bf 62}, 043001 (2000).

\bibitem{Ando02}
S. Ando, K. Sato, and T. Totani,
Astropart. Phys. {\bf 18}, 307 (2003).

\bibitem{Malek02}
M. Malek et al., Super-Kamiokande collaboration,
Phys. Rev. Lett. {\bf 90}, 061101 (2003).

\bibitem{Zhang88}
W. Zhang et al., 
Phys. Rev. Lett. {\bf 61}, 385 (1988).

\bibitem{Fukugita02}
M. Fukugita and M. Kawasaki,
astro-ph/0204376.

\bibitem{Takahashi02b}
K. Takahashi, K. Sato, H.E. Dalhed, and J.R. Wilson,
astro-ph/0212195.

\bibitem{Schirato02}
R.C. Schirato and G.M. Fuller,
astro-ph/0205390.

\bibitem{Dighe00}
A.S. Dighe and A.Yu. Smirnov,
Phys. Rev. D {\bf 62}, 033007 (2000).

\bibitem{Lunardini01}
C. Lunardini and A.Yu. Smirnov,
Nucl. Phys. B {\bf 616}, 307 (2001).

\bibitem{Takahashi02}
K. Takahashi and K. Sato,
Phys. Rev. D {\bf 66}, 033006 (2002).

\bibitem{Totani98}
T. Totani, K. Sato, H.E. Dalhed, and J.R. Wilson,
Astrophys. J. {\bf 496}, 216 (1998).

\bibitem{Keil02}
M.Th. Keil, G.G. Raffelt, and H.T. Janka,
astro-ph/0208035.

\bibitem{Woosley95}
S.E. Woosley and T.A. Weaver,
Astrophys. J. Suppl. {\bf 101}, 181 (1995).

\bibitem{Dziewonski81}
A.M. Dziewonski and D.L. Anderson,
Phys. Earth. Planet. Inter. {\bf 25}, 297 (1981).

\bibitem{Fukuda02}
S. Fukuda et al., Super-Kamiokande collaboration,
Phys. Lett. B {\bf 539}, 179 (2002).

\bibitem{Ahmad02}
Q.R. Ahmad et al., SNO collaboration,
Phys. Rev. Lett. {\bf 89}, 011301 (2002);
{\bf 89}, 011302 (2002).

\bibitem{Fukuda99}
Y. Fukuda et al., Super-Kamiokande collaboration,
Phys. Rev. Lett. {\bf 82}, 2644 (1999).

\bibitem{Takahashi02c}
K. Takahashi and K. Sato,
hep-ph/0205070.

\bibitem{Lunardini01b}
C. Lunardini and A.Yu. Smirnov,
Phys. Rev. D {\bf 63}, 073009 (2001);
H. Minakata and H. Nunokawa,
Phys. Lett. B {\bf 504}, 301 (2001);
H. Minakata,
Nucl. Phys. B (Proc. Suppl.) {\bf 100}, 237 (2001).

\bibitem{Apollonio99}
M. Apollonio et al.,
Phys. Lett. B {\bf 466}, 415 (1999).

\bibitem{Madau96}
P. Madau, H.C. Ferguson, M.E. Dickinson, M. Giavalisco, C.C. Steidel,
	and A. Fruchter,
Mon. Not. R. Astron. Soc. {\bf 283}, 1388 (1996).

\bibitem{Porciani01}
C. Porciani and P. Madau,
Astrophys. J. {\bf 548}, 522 (2001).

\bibitem{Madau98}
P. Madau, M. della Valle, and N. Panagia,
Mon. Not. R. Astron. Soc. {\bf 297}, 17 (1998).

\bibitem{Suzuki02}
Y. Suzuki,
private communication.

\bibitem{Ando02b}
S. Ando and K. Sato,
Phys. Rev. D {\bf 67}, 023004 (2003).

\end{thebibliography}

\clearpage

\end{document}